\documentclass[conference]{IEEEtran}
\usepackage[x11names,dvipsnames,table]{xcolor}
\def\BibTeX{{\rm B\kern-.05em{\sc i\kern-.025em b}\kern-.08em
    T\kern-.1667em\lower.7ex\hbox{E}\kern-.125emX}}

\usepackage[
    disable,     
    textwidth=.9\marginparwidth,
    colorinlistoftodos,
    prependcaption,
    linecolor=green,backgroundcolor=green!25,bordercolor=green
]{todonotes}

\makeatletter
\if@todonotes@disabled
\else 
\paperwidth=\dimexpr \paperwidth + 5cm\relax
\oddsidemargin=\dimexpr\oddsidemargin + 2cm\relax
\evensidemargin=\dimexpr\evensidemargin + 2cm\relax
\marginparwidth=\dimexpr \marginparwidth + 2.5cm\relax

\define@key{todonotes}{TM}[]{%
    \setkeys{todonotes}{linecolor=orange,backgroundcolor=orange!25,bordercolor=orange}
}%
\define@key{todonotes}{TOW}[]{%
    \setkeys{todonotes}{linecolor=blue,backgroundcolor=blue!25,bordercolor=blue}
}%
\fi 
\makeatother

\usepackage[hidelinks]{hyperref}
\usepackage{url}
\usepackage[capitalise]{cleveref}
\usepackage{times}
\usepackage{xspace}

\newcommand{\tool}{\texttt{CTF Pilot}\xspace}

\begin{document}

\title{GitOps for Capture the Flag Platforms%
}

\author{
    \IEEEauthorblockN{
        Mikkel Bengtson Albrechtsen\IEEEauthorrefmark{1}\IEEEauthorrefmark{2},
        Jacopo Mauro\IEEEauthorrefmark{1},
        Torben Worm\IEEEauthorrefmark{1}
    }
    \IEEEauthorblockA{
        \textit{\IEEEauthorrefmark{1}University of Southern Denmark, Odense, Denmark}\\
        \textit{\IEEEauthorrefmark{2}Brunnerne, Denmark}\\
        \vspace{1mm}
        themikkel@brunnerne.dk,\ 
        mauro@imada.sdu.dk,\ 
        tow@mmmi.sdu.dk
    }
}
\maketitle

\begin{abstract}
In this paper, we present \tool, a GitOps-based framework for the deployment and management of Capture The Flag (CTF) competitions. By leveraging Git repositories as the single source of truth for challenge definitions and infrastructure configurations, \tool enables automated, version-controlled deployments that enhance collaboration among challenge authors and organizers. We detail the design criteria and implementation of \tool and evaluate our approach through a real-world CTF event, demonstrating its cost efficiency and its effectiveness in handling high participant concurrency while ensuring robust isolation and ease of challenge development. Our results indicate that \tool improves the experience for organizers and participants, and we present the lessons learned, highlighting opportunities for future improvement.
\end{abstract}


\section{Introduction}
\label{sec:introduction}

Capture The Flag (CTF) competitions~\cite{enisa2021ctf} are a popular format for cybersecurity challenges, where participants solve tasks to earn points and capture "flags". These events often require complex infrastructure to host challenges, manage participant access, and ensure fair play. The (possible) high number of participants in CTF events demands robust infrastructure and scalable solutions that can handle concurrent access and peak loads while maintaining isolation between teams to prevent cheating. The consequences of misconfiguration or downtime can be high, as it can lead to a poor experience for participants and organizers alike.

CTF environments are generally composed of multiple services and applications in order to serve the challenges and collect and present the results~\cite{usenix_CTF}. While the services and challenges are typically containerized, setting up CTF environments can be complex and error-prone, especially when scaling to accommodate a large number of participants.
A particularly persistent difficulty lies in the development and deployment of challenges themselves: authors must manually build, package, and publish challenge artifacts, often relying on ad-hoc scripts or inconsistent deployment practices. These fragmented workflows increase the likelihood of configuration drift, and lead to unpredictable runtime behavior.

To address these issues, we advocate for the usage of Infrastructure-as-Code (IaC) and GitOps. By expressing both infrastructure and challenge definitions declaratively and storing them in version-controlled repositories, organizers can ensure reproducibility and automate the provisioning pipeline. GitOps controllers continuously reconcile the live system with the repository state, reducing manual intervention and enabling challenge authors to focus on content creation instead of operational complexity. This shift turns challenge deployment into a structured, reviewable, and reliable process that scales naturally with the size of the event.

In this paper, we present \tool, a GitOps-driven framework that unifies challenge development, deployment, and infrastructure management for large-scale Capture The Flag (CTF) competitions. Central to our contribution is a development workflow that integrates IaC and GitOps principles to simplify challenge creation, standardize deployment, and ensure consistency across the entire system. The solution supports efficient challenge development, high participant concurrency with strict isolation guarantees. We demonstrate the effectiveness of our approach through its deployment in a large CTF event, presenting participant feedback and discussing the key lessons learned from the operational experience.

The paper is structured as follows: Section \ref{sec:preliminaries} provides background information on CTF competitions and GitOps principles. Section \ref{sec:design_implementation} details the design and implementation of \tool. Section \ref{sec:evaluation} presents the evaluation and discussion of the approach through a real-world CTF event. Section \ref{sec:related} discusses related work in the field. Finally, Section \ref{sec:conclusion} concludes the paper and outlines future work.

\section{Preliminaries}
\label{sec:preliminaries}
In this section, we provide background information starting with basic information on Capture The Flag competitions. 
Next, we introduce the automation paradigms that enable automated, scalable, and reproducible deployment workflows. 
Finally, we describe the key platforms and tools we used for developing \tool.

\subsection{Capture The Flag Competitions}
Capture The Flag (CTF) competitions are cybersecurity exercises in which participants solve challenges to obtain ``flags'', i.e., secret tokens that prove successful exploitation or analysis. CTFs serve a dual purpose: they are widely used in educational settings to foster hands-on learning in security topics, and they form the basis of competitive events at both national and international levels~\cite{enisa2021ctf}.

CTF events typically follow one of two major formats: \emph{Jeopardy-style} and \emph{Attack--Defense}. In Jeopardy-style competitions, participants solve standalone challenges across categories such as web exploitation, cryptography, forensics, mobile security, reverse engineering, and binary exploitation. Each challenge provides a flag upon successful completion and contributes points to the team's score. Attack--Defense competitions, on the other hand, involve teams defending their own vulnerable services while simultaneously attacking the services of others. Although more complex to organize, Attack--Defense CTFs closely mirror real-world adversarial conditions.
In this work, we focus on Jeopardy-style CTFs, which are more prevalent in educational contexts and easier to standardize for automated deployment.


\subsection{Automation Paradigms}

\subsubsection*{Infrastructure as Code}

Infrastructure as Code (IaC) is a paradigm for provisioning and managing computing environments through declarative, machine-readable configuration files rather than manual administrative procedures. 
By treating infrastructure definitions as software artifacts, IaC brings software engineering principle such as version control, automated testing, code review into the domain of system and platform management. 
This approach helps eliminate configuration drift, making system behavior more predictable and transparent~\cite{IaC}.

Modern IaC tools (e.g., OpenTofu/Terraform, Pulumi) support a broad set of cloud and on-premises platforms. 
They provide mechanisms for modularization, dependency resolution, dry-run evaluation, and state inspection, enabling teams to manage complex infrastructures at scale. 

\subsubsection*{Continuous Integration and Delivery}

Continuous Integration and Continuous Delivery (CI/CD) have become foundational practices in modern software engineering. 
The original concept of continuous integration dates back to the late 1980s with early build management systems~\cite{DBLP:conf/compsac/KaiserPS89}. 
Over time, this evolved into full CI/CD pipelines that automate not only build and test but also deployment~\cite{humble2010continuous}.

At their core, CI/CD pipelines orchestrate a series of automated stages. 
Following lightweight pre-commit checks, code is integrated and subjected to developer test suites, static analysis, and integration testing. 
These validations are crucial for detecting regressions and enforcing secure coding practices~\cite{bass2015devops}. 
Compilation then produces distributable artifacts, which can be containerized to guarantee consistent execution across heterogeneous environments. 
The delivery stage promotes artifacts to staging, where both functional and non-functional requirements
are assessed before deployment. 

The abundance of available CI/CD tools reflects the maturity of the DevOps ecosystem. 
Selection often depends on compatibility, vendor support, cost, and platform constraints~\cite{cncf_landscape}. 
Because organizational needs evolve, migrations between tools are frequent. 
Vendor lock-in remains a persistent concern, as pipelines tightly coupled to a single system increase future migration costs. 
To mitigate this, open standards such as the Open Container Initiative~\cite{oci} and containerization strategies have been proposed. 
Security considerations further complicate pipeline design. 
The DevSecOps paradigm emphases embedding security throughout the software lifecycle, 
a necessity in light of widespread supply chain attacks~\cite{supply_chain,bird2020devsecops}. 

\subsubsection*{GitOps}

GitOps~\cite{gitops} is a software operations paradigm that extends the principles of Continuous Delivery into infrastructure management. 
In a GitOps workflow, the entire system configuration—including application manifests, infrastructure definitions, and security policies—is stored as versioned code in a Git repository. 
The state of the repository represents the desired state of the system, while automated agents continuously reconcile the live infrastructure against that repository.

In practice, GitOps merges operational automation with software engineering discipline. 
Changes to manifests trigger automated deployments without manual access to production clusters, reducing configuration drift and operational risk. 
Rollback mechanisms are simplified since any previous system version can be restored by reverting a Git commit. 
The declarative model also facilitates compliance, as every change is reviewable and logged.

\subsection{Platforms and Tooling}

\subsubsection*{Kubernetes}

Kubernetes~\cite{kubernetes} is an open-source container orchestration platform originally developed by Google and later donated to the Cloud Native Computing Foundation (CNCF), becoming the de facto standard for container orchestration in cloud-native environments.
It provides a declarative model for deploying, scaling, and managing containerized applications across clusters of physical or virtual machines. 

At its core, Kubernetes abstracts computing resources as a cluster of \emph{nodes}, each hosting one or more \emph{pods}—the smallest deployable unit in the Kubernetes model.
Pods encapsulate one or more containers that share storage volumes and network interfaces. 
The system’s declarative configuration is defined through YAML manifests that specify the desired cluster state. 
Kubernetes’ control plane continuously reconciles this desired state against the actual state, automatically performing scheduling, restarts, or rescaling as needed. 
Because of its declarative nature and strong automation capabilities, Kubernetes provides a robust substrate for systems that require scalability, self-healing, and multi-tenant isolation. 

\subsubsection*{CTFd}

CTFd~\cite{ctfd} is an open-source web platform for hosting and managing CTF competitions. It implements a complete competition workflow including user registration, team management, challenge deployment, flag submission, scoring, and live leaderboard updates. The system follows a modular architecture written in Python and based on the Flask web framework. Persistent data are stored in a relational database (commonly PostgreSQL or MySQL), while transient state and caching are handled through a Redis backend.

CTFd exposes a RESTful API and a plugin system that allows organizers to extend its functionality—such as custom authentication providers, automated scoring engines, and integrations with orchestration systems. Organizers can create challenges through a graphical interface or by importing structured JSON manifests. Each challenge definition includes metadata (category, points, description, and flag type) and an optional connection string that points to the running instance of the challenge.
However, CTFd by itself does not provide the capability to launch or manage per-team isolated environments. Traditionally, competitions relying solely on CTFd must pre-deploy shared challenges or integrate external automation scripts.

\subsubsection*{kube-ctf}

kube-ctf~\cite{DownUnderCTFKubectf2025} is a instancing system by DownUnderCTF, built to use Kubernetes-native principles to manage CTF challenge instances in an automated and scalable manner.
It extends the Kubernetes control plane by introducing custom resources representing challenges, which it uses to generate per-team deployments. 
When a team deploys and instance, kube-ctf automatically deploys its resources with a team-specific ID to Kubernetes.
Kubernetes scheduling is then leveraged to start and maintain the specified resources.

kube-ctf uses YAML definitions to define the resources to deploy for each team (e.g., deployments, services and ingress)
and network policies to isolate each pod in the instancing namespace. To limit the overall resource usage, there is a preset limit of instances running per team. Kube-janitor \cite{hjacobsKubejanitor} is then used to automatically delete instances after a preset amount of time to free up resources.




\section{System Design and Implementation}
\label{sec:design_implementation}

In this section, we present the overall design and implementation of \tool starting with the key design goals, architectural choices, and the core infrastructure components developed.

\subsection*{Design Goals and Requirements}

\tool was designed to support a modern, large-scale Capture The Flag (CTF) competition combining high participant concurrency with strict isolation guarantees.
Key design goals included:

\begin{itemize}
  \item \textbf{Scalability:} enable dynamic provisioning of hundreds of challenge instances on demand, while maintaining consistent performance.
  \item \textbf{Reliability and fault tolerance:} automatically recover from component failures, maintaining service continuity during competition peaks.
  \item \textbf{Isolation:} ensure that each team interacts only with its own challenge instance, preventing cross-team interference or leakage.
  \item \textbf{Reproducibility:} manage all resources through Infrastructure-as-Code, ensuring complete reproducibility for debugging and root cause analysis in case of problems.
  \item \textbf{Cost efficiency:} maintain low operational expenses, allowing the quick creation and tearing down of the entire infrastructure.
  \item \textbf{Ease of operation:} provide a lightweight management and deployment pipeline for challenge authors and event organizers.
\end{itemize}

\subsection*{Architectural Design and Implementation}

To meet the design goals without reinventing existing solutions, we adopted a hybrid approach that integrates mature, open-source CTF frameworks with custom-built automation and orchestration components. Specifically, we utilized \texttt{CTFd} for the management of challenges and scoreboards, and \texttt{kube-ctf} for dynamic challenge instancing. These systems were extended with our own infrastructure automation layer, enabling seamless interaction between services and allowing fully automated deployment, configuration, and teardown of the CTF environment through declarative configuration files.

\begin{figure*}[t]
    \centering
    \includegraphics[width=\linewidth]{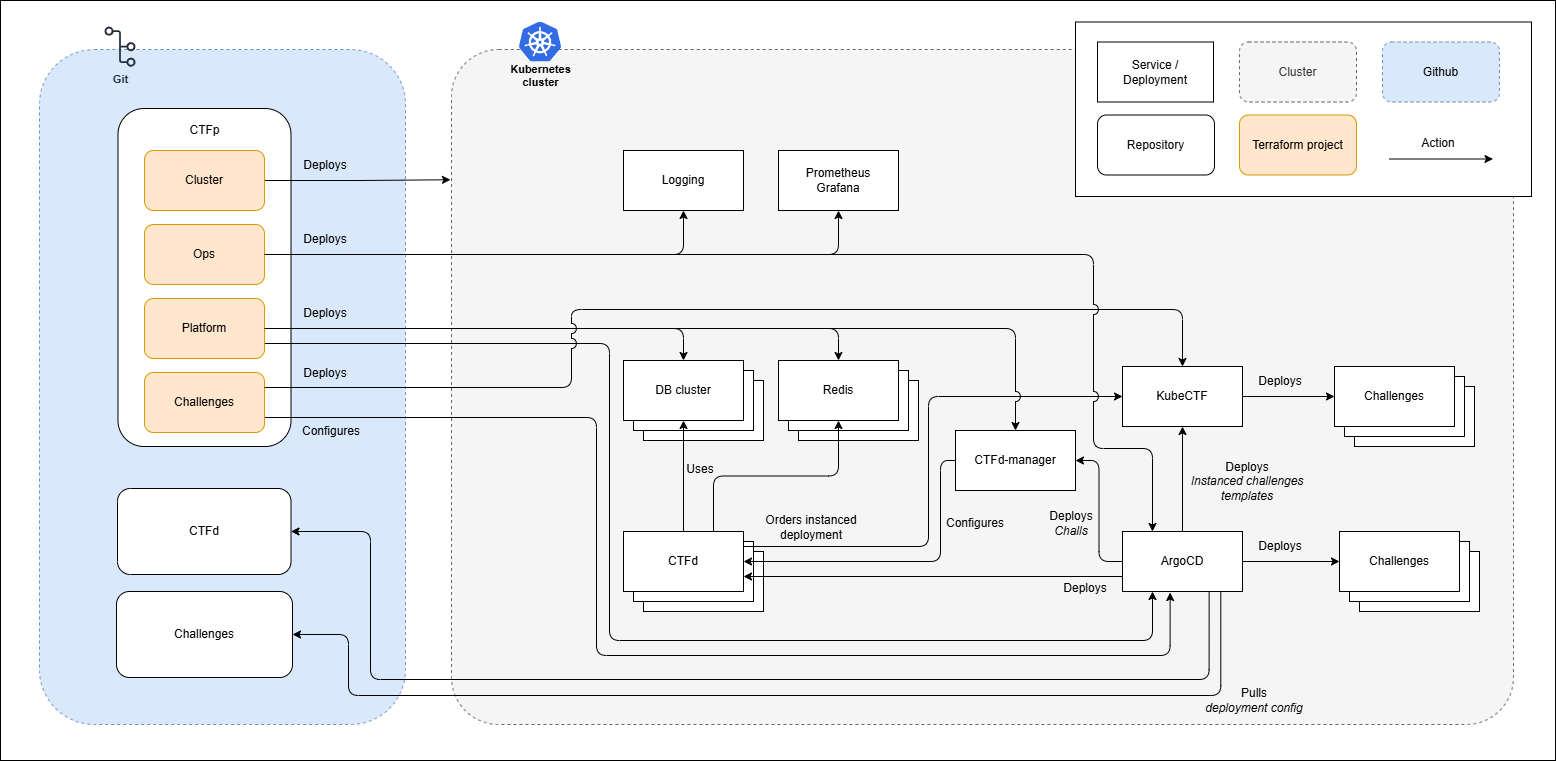}
    \caption{High-level overview of the \tool infrastructure.}
    \label{fig:deployment}
\end{figure*}

This architecture combines the reliability of well-tested CTF management systems with the flexibility of Infrastructure-as-Code, making it possible to reproduce and redeploy the entire environment in a consistent and transparent manner. Figure~\cref{fig:deployment} illustrates the overall system architecture, showing the integration of the major components and their deployment flow within the Kubernetes-based platform.

Kubernetes was selected as the container orchestration engine to host the challenges and all the services due to its ability to ensure fault tolerance and scalability across multiple nodes. The cluster architecture consists of four node pools:
\begin{itemize}
    \item \textbf{Control Planes:} managing the Kubernetes control plane and orchestrating the overall cluster.
    \item \textbf{Agents:} executing general workloads, including monitoring, deployment engines, and the main CTF services.
    \item \textbf{Challs:} dedicated to hosting challenge containers, both shared and instanced.
    \item \textbf{Scale:} an autoscaled pool managed by the Cluster Autoscaler to dynamically allocate additional resources when scheduling constraints occur.
\end{itemize}

\subsection*{Core Infrastructure Components}

Traefik~\cite{traefik} was used as reverse proxy and ingress controller. Traefik surpasses traditional alternative solutions like
Nginx and HAProxy by providing comprehensive support for
various configuration methods, sensible defaults, and features
such as TLS certificate management and observability tools~\cite{rathiPerformanceAnalysisDifferent2024}.
It automatically detects and configures routes to backend services within Kubernetes, supporting dynamic service discovery, SSL termination, and load balancing. Its seamless integration with Kubernetes ingress resources and support for middleware mechanisms make it particularly suited for multi-service deployments such as CTF platforms, where flexibility and automation are essential.

For metrics and performance monitoring, we deployed the \texttt{Prometheus}–\texttt{Grafana} stack~\cite{prometheus, grafana}. \texttt{Prometheus} collects metrics via HTTP-based exporters and provides a powerful query language (PromQL) for real-time analysis. \texttt{Grafana} complements Prometheus by offering a rich, interactive visualization layer, supporting customizable dashboards, alerting mechanisms, and correlation of multiple data sources. Together, they enable precise observability over resource utilization, application performance, and cluster health.

Filebeat~\cite{filebeat} is used to collect logs from containerized applications and system components, and forwarding them to centralized logging solutions. In our setup, it aggregates logs from cluster nodes and application pods, ensuring uniform log collection across distributed components. These logs are then processed and indexed by an external \texttt{Elastic Search}-\texttt{Kibana} stack~\cite{elk}, allowing for long-term storage, advanced querying, and visualization of historical logs.

We developed several custom visualization dashboards to enhance observability and analysis. Specifically, we designed a fully custom \texttt{Grafana} dashboard inspired by~\cite{CTFdExporter}, providing tailored insights into challenge activity and system performance. We also integrated \texttt{Kibana} for log analysis and trend monitoring, creating a dedicated dashboard to track participant behavior and detect the use of automated tools. We complemented these with off-the-shelf \texttt{Grafana} dashboards primarily focused on Kubernetes resource utilization, ensuring a comprehensive monitoring setup spanning both application-level and infrastructure-level metrics.


The network architecture of \tool was designed to ensure both security and scalability by segmenting services into distinct domains. The infrastructure is divided into three primary categories: {Management}, {Challenges}, and {Scoreboard}, each hosted on separate DNS entries. This separation facilitates clearer network policies and reduces the risk of cross-service interference. TLS certificates are managed automatically using the free Let's Encrypt provider~\cite{LetsEncrypt}, enabling seamless renewal and minimizing manual intervention. 
Challenge instances are exposed via wildcard domains to allow dynamic allocation of per-team endpoints without requiring manual DNS configuration. To further enhance security and reliability, we used the free tier of Cloudflare as a protective and caching layer in front of both the Management and Scoreboard services. This setup mitigates Distributed Denial-of-Service (DDoS) risks~\footnote{Previous experience from other CTF organizing teams has experienced them. See, e.g., \cite{InfraWriteup2022}}, provides rate limiting and Web Application Firewall (WAF) features, and improves content delivery performance through its global CDN.
Internally, the cluster nodes are connected via a secure VPN overlay, ensuring encrypted communication between components across physical and virtual boundaries.

\subsection*{Challenges Frontend and Execution}

For scoreboard and challenge frontend, \texttt{CTFd} was deployed within the Kubernetes cluster. The deployment relied on a clustered database and caching layer. Specifically, a multi-node MariaDB Galera cluster was provisioned through the \texttt{mariadb-operator}~\cite{MariadboperatorMariadboperator2025}, while a clustered Redis cache was managed by the \texttt{redis-operator}~\cite{OTCONTAINERKITRedisoperator2025}. These configurations ensured automatic failover and high availability. To ensure robustness, the system performed automated backups every 15 minutes, distributed across multiple physical locations to enhance data resilience. Handout files are served through S3-compatible object storage~\cite{S3StorageSolution}, offloading download traffic from the main infrastructure.

For the deployment and execution of the challenges we got inspired by existing open-source initiatives, particularly DownUnderCTF’s \texttt{kube-ctf}. 
The kube-ctf platform utilize Custom Resource Definitions (CRDs) to store templates of challenges. When deploying a challenge, an API request is sent to the kube-ctf challenge manager, which fills out the templating in the challenge templates, and deploys them to Kubernetes. Thereby relying on the native orchestration layer to manage lifecycle and resource allocation.
We utilized specifically the "challenge-manager", "landing" services and their modified CTFd Plugin that we further change to better personalize the display of the challenges and the possibility to have multiple endpoints to reach the challenge.

Each challenge instance comprised one or more deployments, services, and ingresses, depending on the challenge requirements. To manage states such as “offline” and “loading,” we leveraged \texttt{Traefik}'s routing capabilities. \texttt{kube-ctf} included a fallback page for offline challenges, while custom error middleware was used to indicate “challenge starting” states, enhancing transparency for participants.

\subsection*{Deployment}

For the deployment of \tool, we utilized OpenTofu~\cite{OpenTofu}, an open-source Infrastructure-as-Code tool forked from Terraform, that allows for declarative management of cloud and on-premises resources. The infrastructure deployment was organized into four OpenTofu projects—\textit{Cluster}, \textit{Ops}, \textit{Platform}, and \textit{Challenges}:
\begin{itemize}
    \item \textbf{Cluster:} cluster creation and configuration;
    \item \textbf{Ops:} logging, monitoring, operators, and \texttt{ArgoCD};
    \item \textbf{Platform:} \texttt{CTFd} and supporting services;
    \item \textbf{Challenges:} deployment and management of challenge instances.
\end{itemize}
The use of multiple projects was due to the need to isolate Custom Resource Definitions (CRDs) and minimize state complexity.
Full deployment could be completed in under 30 minutes, and shutdown in under 10 minutes.

In order to facilitate deployment of challenges, we use ArgoCD~\cite{ArgoCD}, i.e., a GitOps tool that synchronize deployment files from Git into Kubernetes. The challenges were split into two deployment files: Challenge Information and Challenge Deployment.
The Challenge Information is a simple configuration file that contains details such as the challenge’s name, description, and structure.
The Challenge Deployment defines how the challenge is launched in the system. Depending on the type of challenge, this could either be a shared setup (using a helm chart) or an individual setup (using a kube-ctf resource to create separate environments for each participant).

To import the challenge information into CTFd, a CTFd manager was implemented as a Go-based service that detects changes of Challenge Informations.
By using the go-ctfd library~\cite{CtferioGoctfd2025}, it synchronizes the Challenge Information into CTFd
and watches for new and updated challenge information continuosly updating CTFd to match the expected state.
As far as Challenge Deployments are concerned, these were automatically handled by ArgoCD, with kube-ctf handling per-team instances for instanced deployments. 

\subsection{Facilitating Challenge Development}

\begin{figure*}[t]
    \centering
   \includegraphics[width=0.6\linewidth]{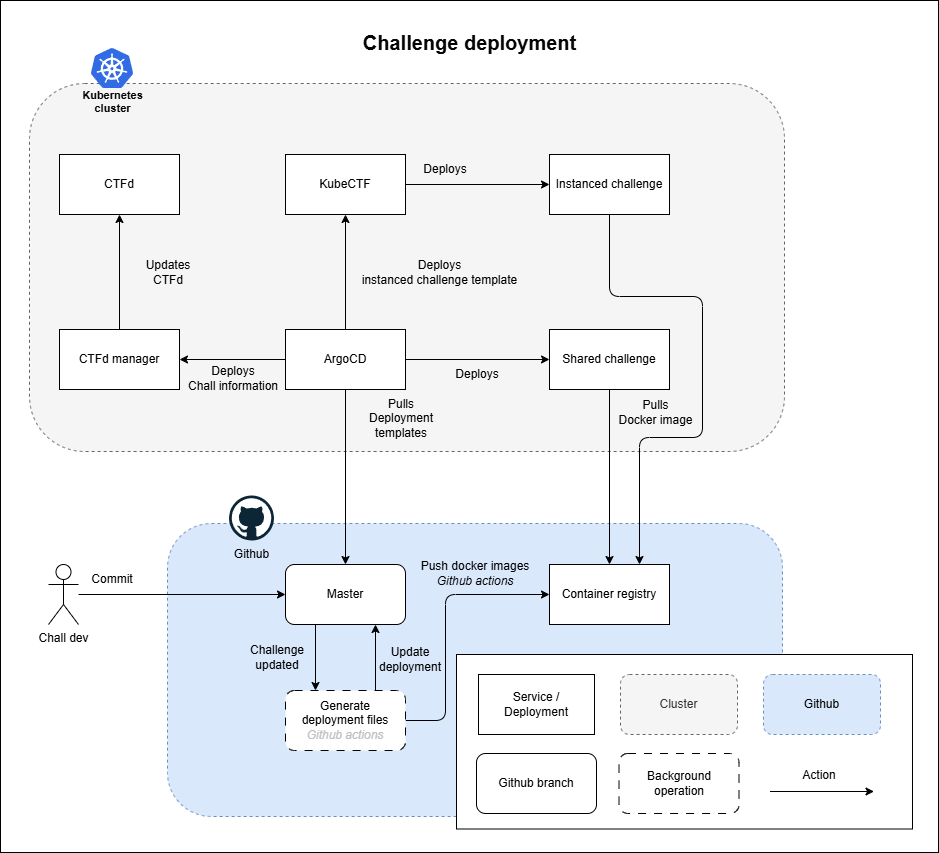}
    \caption{Challenge development workflow.}
    \label{fig:challenges}
\end{figure*}

To streamline the challenge development process, a series of tools were introduced to promote standardization, automation, and traceability. The workflow, depicted in \cref{fig:challenges} begins with Discord~\cite{Discord}, where a custom bot integrates challenge discussion and management directly within the communication platform. The bot automates project initialization by creating a GitHub issue, a dedicated branch, and an associated pull request (PR), effectively linking challenge ideation with version-controlled development. The GitHub issue provides progress tracking through a project board, while the PR enables collaborative review and integration.

Each challenge adheres to a standardized schema describing metadata such as name, type, flags, associated Dockerfiles, and scoring information. This schema ensures consistency and enables automated tooling to interact programmatically with challenge definitions. Based on this structure, challenge builds were fully automated. Docker images were generated and published to the GitHub Container Registry, while handout materials placed in designated directories were automatically packaged into ZIP archives for participant distribution.

This automated pipeline allowed developers to focus solely on challenge logic and design, abstracting away the complexity of packaging, deployment, and distribution. The resulting system substantially reduced manual overhead and improved reproducibility across the challenge lifecycle.

All source code developed for this work is available at https://github.com/ctfpilot.
In total, the codebase comprises approximately
3.8k lines of Python, 2.8k lines of Go, and 8.2k lines for OpenTofu.

\section{Evaluation and Discussion}
\label{sec:evaluation}    

The scalability, reliability, and user experience of the developed CTF platform was assessed in a large-scale CTF competition, called BrunnerCTF 2025 \cite{teamBrunnerCTF2025}. 
The event had registered a total of 2,860 participants and 1,491 teams. Among these, 1,158 teams obtained at least the minimum score of 10 points, demonstrating a wide level of engagement.

The CTF featured 83 challenges across 10 categories: \emph{Shake \& Bake for beginners} (21), \emph{Boot2Root} (6), \emph{Crypto} (9), \emph{Forensics} (5), \emph{Misc} (10), \emph{Mobile} (3), \emph{OSINT} (4), \emph{Pwn} (6), \emph{Reverse Engineering} (9), and \emph{Web} (10). All challenges were solved at least once, with the winning team solving 79 out of 83 challenges.
The platform supported three types of challenges: \emph{Static} (54), \emph{Shared} (6), and \emph{Instanced} (23). Static challenges required only textual descriptions or downloadable files, Shared challenges ran on a common remote host accessible to all participants, and Instanced challenges provisioned isolated environments per team. At peak load, approximately 600 instances were simultaneously active, consuming around 8 vCPUs and 43~GB of RAM. During the first hour, roughly 750 instances were deployed.

The Kubernetes-based infrastructure was hosted on {Hetzner} \cite{AffordableDedicatedServers},
chosen for their favorable cost-performance ratio and the fact that their servers were located in Europe. The cluster comprised 164 vCPUs and 386 GB of RAM distributed across 14 nodes: three for the control planes (CPX31 type), three for the {agent nodes} (CCX33 type), seven for the {challenge nodes} (CPX51 type), and one for the scaling node (CPX51 type). Peak resource utilization reached approximately 26 vCPUs and 67 GB RAM, leaving significant unused capacity and demonstrating robust headroom for load fluctuations.
Deployment automation was achieved using \emph{Kube-Hetzner} \cite{KubehetznerTerraformhcloudkubehetzner2025a}, an open-source Terraform-based framework for Kubernetes provisioning on Hetzner servers. This setup facilitated rapid deployment, reproducibility, and configuration transparency. Despite the scale of the infrastructure, costs remained remarkably low: the total expenditure, including both development and production clusters, amounted to 265.18~€. The production environment alone, active from early August until the event’s conclusion on August~25, cost 127.45~€. The development cluster, maintained for approximately 1.5~months to enable testing and challenge validation, accounted for the remaining 137.73~€. These figures demonstrate that high-performance CTF infrastructure can be achieved at minimal expense through careful resource management and the use of cost-effective bare-metal providers.

Network metrics indicated between 1,500 and 2,500 concurrent connections at any given time. Over the course of the event, the platform processed more than 35 million HTTP requests, with peak throughput around 160,000 requests per minute. Some instances received over one million requests individually and CTFd accounted for approximately 3 million requests as reported by Cloudflare. Analysis of incoming traffic through user agents provided by HTTP traffic logging revealed numerous brute-force attempts, highlighting the need for effective rate limiting and monitoring mechanisms.

\subsection*{Operational Issues and Mitigation}

The event experienced a partial outage during the first 22 minutes. Initially suspected to be caused by CTFd, the issue was eventually traced to the \texttt{Traefik} ingress controller. Due to a misconfiguration in the autoscaling setup, caused by missing resource specifications for a log-forwarding sidecar, the controller failed to scale beyond its minimum of three replicas. The resulting overload led to slow responses and repeated container restarts. Temporary mitigation involved manually allocating additional CPU and memory resources, immediately stabilizing the platform. Once proper resource limits were defined, CPU utilization across the cluster dropped by over 90\%, and performance normalized.

Although CTFd did not directly cause downtime, the use of a clustered Redis cache introduced stability issues, as CTFd does not natively support this configuration. The system experienced intermittent crashes due to key retrieval errors within the Flask caching library. Redundancy via six concurrent replicas mitigated user-facing downtime. For future deployments a non-clustered Redis setup with failover support to improve reliability should be considered.

Email verification was enforced for all participants, resulting in more than 6,000 outbound messages. Using the Brevo email service \cite{BrevoEmailMarketing} ensured high deliverability. 
However, Google temporarily deferred emails due to domain-based rate limiting. The issue was mitigated by switching sender domains mid-event. For future events, multiple preconfigured domains should be considered to prevent throttling.

\subsection*{Lessons Learned}

Post-event feedback from participants was overwhelmingly positive, particularly regarding the quality of challenges, system stability, and the responsiveness of the organizing team. This outcome suggests that the platform successfully balanced scalability, reliability, and user experience, validating its suitability for future educational and competitive cybersecurity events. The infrastructure was also highlighted in reviews of the CTF, with participating highligting that ``The quality of challenges + infra was insanely high, the challenges also had immense variation in difficulty, from beginner to advanced''~\cite{teamBrunnerCTF2025}.
A participant feedback survey with 39 respondant found that participants where satisfied with the CTF and infrastructure (85\% gave it a score of 4 or 5 on a 5 Likert scale). Respondents in particular appreciated small details like the loading screen on instance startup done by exploiting Traefik status middleware serving ``No service available'' status in the loading page.

Although the challenge development workflow was largely automated, several pain points were identified throughout deployment and evaluation. One recurring limitation was the inability to test challenges directly on the cluster before full deployment. This restriction occasionally led to late-stage reengineering and adjustments under time pressure. Introducing a dedicated pre-deployment testing environment—capable of instantiating challenges in an isolated, production-like setting—would significantly streamline validation and reduce the risk of deployment-time failures. Additionally, large challenge assets caused repository management difficulties, underscoring the need for external storage solutions or content delivery mechanisms to minimize repository bloat and improve scalability for future editions.

A survey run after the CTF challenge with 9 CTF authors highlights that they felt overwhelmed in figuring out how to develop and review challenges. Moreover, they mentioned that it was difficult to have an overview over what challenges needed review, and navigate the challenges when the repository grew to a large challenge count.
Several improvements were suggested for future iterations. Some authors proposed organizing a dedicated workshop to introduce new contributors to the system, noting that several participants had limited experience with GitHub and had not previously used GitHub Projects or Issues to track progress. Others recommended providing a clear overview of the files to modify in the base template, to help authors understand which parts to edit and which to leave unchanged.



Empirical evidence from the event further confirmed the operational advantages of instanced environments. During the competition, shared challenge boxes experienced performance degradation due to overloading and CPU throttling, particularly in the case of TCP-based challenges where limited logging capabilities prevented accurate attribution of excessive load to specific teams. In contrast, instanced challenges operated in isolated containers, ensuring that each team’s resource consumption affected only their own environment. This approach not only maintained consistent performance across participants but also improved reliability and facilitated incident tracking. These observations highlight that large-scale instancing is both technically feasible and essential for preserving fairness, stability, and observability in competitive cybersecurity environments.

Monitoring and observability played a critical role in maintaining platform stability. While Prometheus and Grafana offered robust real-time dashboards, the absence of fine-grained alert rules limited proactive issue detection. Certain performance anomalies—such as CPU throttling, overloaded remotes, and excessive network utilization—were identified only after prolonged observation. Enhancing the monitoring stack with tailored alert policies,
and automated anomaly detection would improve operational responsiveness. 

As far as security is concerned, despite clear rules prohibiting scanning and brute-force attacks, participant behavior repeatedly demonstrated that such activities occur regardless of formal restrictions. Consequently, future iterations of the platform must incorporate stronger safeguards, such as rate limiting, sandboxing, and automated intrusion detection. Integrating these security controls into the logging and monitoring framework would not only protect system integrity but also provide valuable insights into emerging attack vectors.

In our deployment, sandboxing was implemented to the extent possible within the constraints of Kubernetes. We did not observe significant issues related to participants attempting to abuse the platform, as each containerized challenge instance was constrained by CPU limits, effectively providing a natural form of rate limiting. There was a single attempt to escape a container; however, it was unsuccessful. Discussions with others CTF organizing groups revealed that container escape remains their primary concern.

As far as security is concerned, we would like to note that due to the adoption of the GitOps approach our entire infrastructure is automatically updated from Git repositories. To do so we store challenge details inside the Kubernetes cluster and an attacker who gains access to the system could potentially see sensitive information like challenge data (e.g., their source code) or important credentials (e.g., database password and administrator access token for CTFd). 
One solution to securing the infrastructure, would be to split the infrastructure in two separate clusters, one responsible for the scoreboard and management and one to deploy the challenge instances. This minimizes the possibility for lateral movements within the infrastructure.
Clearly, in this case, more resources would be needed to operate a second cluster.



\subsection*{Vendor Lock-In Considerations}

The current platform relies on a few external non-open source dependencies or providers, most notably {Cloudflare}, Traefik and {GitHub}. The use of {Cloudflare} and Traefik primarily serves as a content delivery network and reverse proxy. This dependency is non-critical and could easily be replaced with alternative CDN providers without substantial architectural changes, as the integration is limited to standard DNS and caching configurations.

In contrast, the reliance on \texttt{GitHub} is more deeply integrated into the platform’s workflow. Several components, including continuous integration pipelines and project management, are implemented using \texttt{GitHub Actions} and GitHub’s project boards. These integrations are vendor-specific and would not directly transfer to other Git-based services (e.g., GitLab) without modification. While the codebase itself could be migrated with relative ease, reproducing the automation and coordination mechanisms would require reimplementation using the respective platform’s CI/CD and project management tools. Thus, although the GitHub dependency is not fundamental to the design, it remains a practical constraint due to the convenience and maturity of the GitHub ecosystem.

\section{Related Work}
\label{sec:related}

A number of platforms have been developed to support the deployment and management of Capture the Flag (CTF) competitions. These systems vary in their degree of automation, deployment strategies, and integration with scoring platforms.

Commercial platforms, such as Parrot CTFs \cite{teamParrotCTFsEvents}, Hack The Box \cite{boxAttackExecution}, TryHackMe CTF Builder \cite{IntroducingOurNew}, and CTFd Enterprise \cite{ctfd}, offer comprehensive ecosystems that combine challenge deployment, scoring, and user management within managed infrastructures. While these services significantly reduce the operational overhead for organizers, they often come at the expense of transparency and flexibility, depending on vendor-maintained infrastructures and closed-source implementations.

Open-source initiatives such as kCTF \cite{GoogleKctf2025} and kube-ctf \cite{DownUnderCTFKubectf2025} take a different approach by focusing on scalable deployment and isolation within Kubernetes. Google’s kCTF provides a hardened environment for securely running containerized challenges and has been widely adopted for large-scale competitions. Similarly, kube-ctf introduces Kubernetes-native abstractions to manage per-team challenge instances and automate their lifecycle. Both systems, however, concentrate on the runtime and isolation aspects of CTF deployment, rather than on the full authoring and synchronization pipeline from challenge definition to scoreboard publication.

CTFer.io \cite{CtferioGoctfd2025} represents another relevant development. It adopts a GitOps-based approach similar to ours, offering a SaaS platform for managing challenge definitions and synchronizing them with CTFd. Its released components include a challenge manager and a CTFd integration for the challenge manager. However, CTFer.io currently provides only partial open-source coverage of its components, limiting its adaptability to self-hosted or research-oriented contexts.

Another interesting platform is Haaukins \cite{Haaukins}, which focuses primarily on providing each participant with an isolated laboratory environment. This lab can be accessed either through a virtual private network or via a web-based Linux interface. Haaukins emphasizes educational settings and individualized environments, but  does not specifically address and facilitate challenge creation and testing.

 Compared to these existing systems, \tool aims to bridge the gap between infrastructure management and challenge authoring by adopting a fully GitOps-driven approach. This workflow not only supports auditability and version control but also aligns naturally with standard software development practices (e.g., peer review, automated testing).


\section{Conclusions and Future Work}
\label{sec:conclusion}

This paper presented \tool, a GitOps-driven framework designed to support the development, deployment, and operation of large-scale Jeopardy-style Capture The Flag (CTF) competitions. 
We evaluated the system through its deployment in a real-world CTF event involving thousands of participants and over eighty challenges. 

Overall, the deployment experience confirms that the design choices underlying \tool provide an effective foundation for automating large-scale CTF events, delivering both technical robustness and an improved experience for participants and organizers.
The system successfully scaled to handle hundreds of simultaneous challenge instances with stable performance. 
Isolation guarantees were fulfilled through the use of dedicated instanced deployments, which prevented cross-team interference and facilitated granular observability. 
Reproducibility was achieved through a combination of Infrastructure as Code and GitOps, enabling full cluster creation and teardown within minutes and ensuring transparent versioning of all infrastructure components. 
Operational overhead was significantly reduced through automation layers such as the challenge manager and CI/CD-driven artifact generation, while the hosting strategy kept overall costs remarkably low. 
Finally, the standardized development workflow for the challenges improved consistency and packaging errors, though the evaluation also revealed several areas where authoring could be simplified further.

A central direction for future work is to further simplify the deployment model. This could be achieved by abstracting away low-level deployment details and embedding essential configuration information (e.g., container definitions, exposed services) directly within the challenge description. Such an approach would lower the entry barrier for new contributors while maintaining the ability to define advanced setups involving multiple containers or complex networking requirements.

Another promising avenue for improvement concerns the integration between the infrastructure and the communication tools used during the challenge development process. In particular, the Discord integration could be extended to improve transparency and coordination among contributors. For example, implementing a command that lists challenges currently under review would allow participants to obtain a real-time overview of pending tasks, thereby facilitating collaboration during the review phase.

The GitOps workflow itself also presents opportunities for refinement. One potential enhancement would be to separate automatically generated artifacts (i.e., compiled deployment manifests, derived configuration files) into a distinct repository. This would shield challenge authors from direct exposure to low-level infrastructure code and reduce the likelihood of accidental misconfigurations during review.

\bibliographystyle{IEEEtran}
\bibliography{biblio}

@software{ctfd,
  author       = {{CTFd LLC}},
  title        = {{CTFd: The Easiest Capture the Flag Framework}},
  year         = {2025},
  howpublished = {\url{https://ctfd.io}}
}

@software{DownUnderCTFKubectf2025,
  title        = {{{DownUnderCTF}}/Kube-Ctf},
  date         = {2025-10-27T08:12:53Z},
  origdate     = {2021-09-26T06:16:08Z},
  url          = {https://github.com/DownUnderCTF/kube-ctf},
  urldate      = {2025-10-30},
  organization = {DownUnderCTF}
}

@online{LetsEncrypt,
  author       = {{Internet Security Research Group}},
  title        = {{Let's Encrypt}},
  year         = {2025},
  howpublished = {\url{https://letsencrypt.org}}
}

@misc{MariadboperatorMariadboperator2025,
  author       = {{MariaDB Operator Team}},
  title        = {{MariaDB Operator}},
  year         = {2025},
  howpublished = {\url{https://github.com/mariadb-operator/mariadb-operator}},
}

@misc{OTCONTAINERKITRedisoperator2025,
  author       = {{Opstree Solutions}},
  title        = {{Redis Operator}},
  year         = {2025},
  howpublished = {\url{https://github.com/OT-CONTAINER-KIT/redis-operator}}
}

@inproceedings{DBLP:conf/compsac/KaiserPS89,
  author    = {Gail E. Kaiser and
               Dewayne E. Perry and
               William M. Schell},
  title     = {Infuse: fusing integration test management with change management},
  booktitle = {Proceedings of the 13th Annual International Computer Software and
               Applications Conference, {COMPSAC}},
  pages     = {552--558},
  publisher = {{IEEE}},
  year      = {1989},
  doi       = {10.1109/CMPSAC.1989.65147},
  bibsource = {dblp computer science bibliography, https://dblp.org}
}

@book{humble2010continuous,
  title     = {Continuous delivery: reliable software releases through build, test, and deployment automation},
  author    = {Humble, Jez and Farley, David},
  year      = {2010},
  publisher = {Pearson Education}
}

@book{bass2015devops,
  title     = {DevOps: A Software Architect's Perspective},
  author    = {Bass, Len and Weber, Ingo and Zhu, Liming},
  year      = {2015},
  publisher = {Addison-Wesley Professional},
  isbn      = {9780134049847}
}

@misc{cncf_landscape,
  author       = {{Cloud Native Computing Foundation}},
  title        = {{CNCF Cloud Native Landscape}},
  year         = {2025},
  howpublished = {\url{https://landscape.cncf.io/}}
}

@misc{oci,
  author       = {{Open Container Initiative}},
  title        = {{Open Container Initiative}},
  year         = {2025},
  howpublished = {\url{https://opencontainers.org/}}
}

@book{bird2020devsecops,
  title     = {DevSecOps: A leader's guide to producing secure software without compromising flow, feedback and continuous improvement},
  author    = {Bird, Jim},
  year      = {2020},
  publisher = {Addison-Wesley Professional},
  isbn      = {9780136824855}
}

@article{supply_chain,
  author     = {Cox, Russ},
  title      = {{Fifty Years of Open Source Software Supply Chain Security: For decades, software reuse was only a lofty goal. Now it's very real.}},
  year       = {2025},
  issue_date = {January/February 2025},
  publisher  = {Association for Computing Machinery},
  address    = {New York, NY, USA},
  volume     = {23},
  number     = {1},
  issn       = {1542-7730},
  doi        = {10.1145/3722542},
  journal    = {Queue},
  pages      = {84–107}
}

@software{kubernetes,
  author       = {{The Kubernetes Authors}},
  title        = {Kubernetes: Production-Grade Container Orchestration},
  year         = {2025},
  howpublished = {\url{https://kubernetes.io/}}
}

@misc{gitops,
  author       = {{GitLab Inc.}},
  title        = {{What is GitOps?}},
  year         = {2025},
  howpublished = {\url{https://about.gitlab.com/topics/gitops/}}
}

@misc{traefik,
  title        = {Traefik},
  author       = {{Traefik Labs}},
  year         = {2025},
  howpublished = {\url{https://traefik.io}}
}

@misc{filebeat,
  title        = {Filebeat},
  author       = {{Elastic}},
  year         = {2025},
  howpublished = {\url{https://www.elastic.co/beats/filebeat}}
}

@misc{elk,
  title        = {The Elastic Stack},
  author       = {{Elastic}},
  year         = {2025},
  howpublished = {\url{https://www.elastic.co/elastic-stack}}
}

@misc{prometheus,
  title        = {Prometheus},
  author       = {{Prometheus Authors}},
  year         = {2025},
  howpublished = {\url{https://prometheus.io}}
}

@misc{grafana,
  title        = {Grafana},
  author       = {{Grafana Labs}},
  year         = {2025},
  howpublished = {\url{https://grafana.com}}
}

@inproceedings{rathiPerformanceAnalysisDifferent2024,
  author    = {Rathi, Garima and Amin, Safiya and Jain, Samyak and Yachawad, Mansi and Digholkar, Kirti},
  booktitle = {2024 IEEE International Conference on Information Technology, Electronics and Intelligent Communication Systems (ICITEICS)},
  title     = {{Performance Analysis of Different Ingress Controllers Within the Kubernetes Cluster}},
  year      = {2024},
  volume    = {},
  number    = {},
  pages     = {1-6},
  doi       = {10.1109/ICITEICS61368.2024.10625280}
}

@online{teamBrunnerCTF2025,
  author       = {{Brunnerne}},
  title        = {{{CTFtime}}.Org / {{BrunnerCTF}} 2025},
  howpublished = {\url{https://ctftime.org/event/2835}}
}

@online{CTFdExporter,
  author       = {{Grafana Labs}},
  title        = {{CTFd Exporter}},
  howpublished = {\url{https://grafana.com/grafana/dashboards/23095-ctfd/}},
  urldate      = {2025-10-09}
}

@misc{OpenTofu,
  author       = {{OpenTofu}},
  title        = {{OpenTofu}},
  howpublished = {\url{https://opentofu.org/}}
}

@misc{ArgoCD,
  author       = {{Argo Project}},
  title        = {{Argo CD}},
  howpublished = {\url{https://argoproj.github.io/cd/}}
}

@misc{CtferioGoctfd2025,
  author       = {{CTFer.io}},
  title        = {{go-ctfd}},
  year         = {2025},
  howpublished = {\url{https://github.com/ctfer-io/go-ctfd}}
}

@misc{S3StorageSolution,
  author       = {{Hetzner Online GmbH}},
  title        = {{S3 Storage Solution: Object Storage by Hetzner}},
  howpublished = {\url{https://www.hetzner.com/storage/object-storage/}}
}

@misc{hjacobsKubejanitor,
  author       = {{Kube-Janitor Authors}},
  title        = {{Kube-Janitor}},
  year         = {2025},
  howpublished = {\url{https://codeberg.org/hjacobs/kube-janitor}},
}

@misc{InfraWriteup2022,
  author       = {{DownUnderCTF Team}},
  title        = {{Infra Writeup 2022: DownUnderCTF}},
  year         = {2022},
  howpublished = {\url{https://downunderctf.com/}},
  noteOPT         = {The largest CTF in the Southern Hemisphere}
}

@misc{AffordableDedicatedServers,
  author       = {{Hetzner Online GmbH}},
  title        = {{Affordable Dedicated Servers, Cloud \& Hosting from Germany}},
  howpublished = {\url{https://www.hetzner.com/}}
}

@misc{BrevoEmailMarketing,
  author       = {{Brevo}},
  title        = {{Brevo Email Marketing Software, Automation \& CRM}},
  howpublished = {\url{https://www.brevo.com/}}
}

@misc{boxAttackExecution,
  author       = {{Hack The Box}},
  title        = {{Attack Execution}},
  howpublished = {\url{https://app.arcade.software/share/KfeufQyCnKZFmKdMfrsg}}
}

@online{IntroducingOurNew,
  title        = {Introducing Our {{New CTF Builder Feature}}!},
  url          = {https://tryhackme.com/resources/blog/new-ctf-builder},
  urldate      = {2025-11-04},
  langid       = {english},
  organization = {TryHackMe}
}

@online{teamParrotCTFsEvents,
  title        = {Parrot {{CTFs Events}}: {{Cybersecurity Competitions}} \& {{Workshops}}},
  shorttitle   = {Parrot {{CTFs Events}}},
  url          = {https://parrot-ctfs.com/events},
  urldate      = {2025-11-04},
  langid       = {english},
  organization = {Parrot CTFs}
}

@inproceedings{Haaukins,
  author    = {Thomas Kobber Panum and
               Kaspar Hageman and
               Jens Myrup Pedersen and
               Ren{\'{e}} Rydhof Hansen},
  title     = {{Haaukins: A Highly Accessible and Automated Virtualization Platform
               for Security Education}},
  booktitle = {19th {IEEE} International Conference on Advanced Learning Technologies,
               {ICALT} 2019, Macei{\'{o}}, Brazil, July 15-18, 2019},
  pages     = {236--238},
  publisher = {{IEEE}},
  year      = {2019},
  url       = {https://doi.org/10.1109/ICALT.2019.00073},
  doi       = {10.1109/ICALT.2019.00073},
  bibsource = {dblp computer science bibliography, https://dblp.org}
}

@techreport{enisa2021ctf,
  title        = {{CTF events -- Contemporary practices and state-of-the-art in capture-the-flag competitions}},
  author       = {{European Union Agency for Cybersecurity (ENISA)}},
  year         = {2021},
  institution  = {ENISA},
  howpublished = {\url{https://www.enisa.europa.eu/publications/ctf-events-contemporary-practices-and-state-of-the-art}}
}

@inproceedings{usenix_CTF,
  author       = {Clark Taylor and Pablo Arias and Jim Klopchic and Celeste Matarazzo and Evi Dube},
  title        = {{CTF: State-of-the-Art and Building the Next Generation}},
  booktitle    = {2017 USENIX Workshop on Advances in Security Education (ASE 17)},
  year         = {2017},
  address      = {Vancouver, BC},
  howpublished = {\url{https://www.usenix.org/conference/ase17/workshop-program/presentation/taylor}},
  publisher    = {USENIX Association}
}

@book{IaC,
  author    = {Morris, Kief},
  title     = {Infrastructure as Code: Managing Servers in the Cloud},
  year      = {2016},
  isbn      = {1491924357},
  publisher = {O'Reilly Media, Inc.},
  edition   = {1st}
}

@software{KubehetznerTerraformhcloudkubehetzner2025a,
  author       = {{Kube-Hetzner}},
  title        = {{Terraform Hcloud Kube-Hetzner}},
  year         = {2025},
  howpublished = {\url{https://github.com/kube-hetzner/terraform-hcloud-kube-hetzner}}
}

@software{GoogleKctf2025,
  author       = {{Google}},
  title        = {{kCTF}},
  year         = {2025},
  howpublished = {\url{https://github.com/google/kctf}}
}

@misc{Discord,
  author       = {{Discord Inc.}},
  title        = {{Discord}},
  howpublished = {\url{https://discord.com}}
}

\end{document}